\documentclass[a4paper,11pt]{article}
\usepackage{pos}
\usepackage{subcaption}
\usepackage{braket}
\usepackage[font=small, labelfont=bf, textfont={small,it}]{caption}
\graphicspath{{./}}

\newcommand{\beq}{\begin{eqnarray}}
\newcommand{\eeq}{\end{eqnarray}}
\newcommand{\beqnn}{\begin{eqnarray*}}
\newcommand{\eeqnn}{\end{eqnarray*}}
\newcommand{\Tr}{\mathrm{Tr}}

\newcommand{\SU}{\mathrm{SU}}

\newcommand{\clov}{{\scriptscriptstyle{\mathrm{clov}}}}
\newcommand{\W}{{\scriptscriptstyle{\mathrm{W}}}}
\newcommand{\cool}{{\scriptscriptstyle{\mathrm{cool}}}}
\newcommand{\tor}{{\scriptscriptstyle{\mathrm{tor}}}}
\newcommand{\zpp}{{\scriptscriptstyle{0^{++}}}}
\newcommand{\G}{{\scriptscriptstyle{\mathrm{G}}}}
\renewcommand{\L}{{\scriptscriptstyle{\mathrm{L}}}}
\newcommand{\I}{{\scriptscriptstyle{\mathrm{I}}}}

\title{The imaginary-$\theta$ dependence of the SU($N$) spectrum}
\ShortTitle{The imaginary-$\theta$ dependence of the SU($N$) spectrum}

\author[a]{Claudio Bonanno}
\author[b]{Claudio Bonati}
\author[c]{Mario Papace}
\author*[d]{Davide Vadacchino}
\affiliation[a]{Instituto de F\'isica T\'eorica UAM-CSIC, c/ Nicol\'as Cabrera 13-15, Universidad Aut\'onoma de Madrid, Cantoblanco, E-28049 Madrid, Spain}
\affiliation[b]{Dipartimento di Fisica dell'Universit\`a di Pisa and INFN - Sezione di Pisa, Largo Bruno Pontecorvo 3, I-56127 Pisa, Italy}
\affiliation[c]{Fachbereich Mathematik und Naturwissenschaften, Bergische Universität Wuppertal, Gaußstraße 20, 42119 Wuppertal, Germany}
\affiliation[d]{Centre for Mathematical Sciences, University of Plymouth, Plymouth, PL4 8AA, United Kingdom}

\emailAdd{claudio.bonanno@csic.es}
\emailAdd{claudio.bonati@unipi.it}
\emailAdd{mario.papace@uni-wuppertal.de}
\emailAdd{davide.vadacchino@plymouth.ac.uk}

\abstract{In this talk we will report on a study of the $\theta$-dependence of the string tension and of the mass gap of four-dimensional SU($N$) Yang--Mills theories. The spectrum at $N=3$ and $N=6$ was obtained on the lattice at various imaginary values of the $\theta$-parameter, using Parallel Tempering on Boundary Conditions to avoid topological freezing at fine lattice spacings. The coefficient of the $\mathcal{O}(\theta^2)$ term in the Taylor expansion of the spectrum around $\theta=0$ could be obtained in the continuum limit for $N=3$, and on two fairly fine lattices for $N=6$.}

\FullConference{The 41$^{\text{st}}$ International Symposium on Lattice Field Theory (LATTICE2024)\\
 28 July -- 3 August 2024\\
Liverpool, UK\\}

\begin{document}
\maketitle

\section{Introduction}
\vspace*{-2\baselineskip}
The study of systems described by a Yang-Mills (YM) action augmented with
a $\theta$-term has provided, historically, several interesting results.
Classically speaking, the addition of a $\theta$-term does not alter the classical
equations of motion. 
Yet, its theoretical and phenomenological implications are far reaching, once
the quantum fluctuations are taken into account.
The $\theta$-dependence lies at the heart of the Witten--Veneziano solution
of the $\eta$-$\eta'$ puzzle~\cite{Witten:1979vv,
Veneziano:1979ec,Kawarabayashi:1980dp,Witten:1980sp} and it is the starting
point for the Peccei--Quinn
hypothetical solution to the strong-CP problem based on axions. Accordingly,
the study of $\theta$-dependence in a variety of different settings has 
attracted a great deal of attention in the past years. Starting from QCD, SU($N$)
and Sp($2N$) models~\cite{Coleman:1985rnk,
Gross:1980br,Schafer:1996wv} in 
$4d$, to $CP^{N-1}$ models~\cite{DAdda:1978vbw,Witten:1978bc} and U(N)
models~\cite{Kovacs:1995nn,Bonati:2019ylr,Bonati:2019olo} in $2d$, to quantum
mechanical models~\cite{Jackiw:1979ur,Gaiotto:2017yup,
Bonati:2017woi}.

So far, the main target of investigation has been the free energy of the system
and its $\theta$-dependence, both at zero and finite temperature. Using analytical methods, it is possible to obtain theoretical predictions for
the $\theta$-dependence of the QCD vacuum energy either using effective theories
close to the chiral limits for low temperatures~\cite{DiVecchia:1980yfw,
diCortona:2015ldu,Guo:2015oxa,Lu:2020rhp}, where the $\theta$-dependence of the vacuum energy essentially stems from the $\theta$-dependence of the pion mass (the $\theta$-dependence of the mass of some light resonances has also been investigated in~\cite{Acharya:2015pya}); or for the $\theta$-dependence of the QCD free energy via
perturbative and semi-classical arguments for asymptotically-high
temperatures~\cite{Gross:1980br,Schafer:1996wv,Boccaletti:2020mxu}.
Away from
these regimes, the $\theta$-dependence of the vacuum energy has been
investigated by means of numerical Monte Carlo simulations of the
lattice-discretized theory, both in QCD~\cite{Bonati:2015vqz,
Petreczky:2016vrs,Frison:2016vuc,Borsanyi:2016ksw,Bonati:2018blm,Burger:2018fvb,Chen:2022fid,Athenodorou:2022aay}
and in $\SU(N)$ pure-gauge
theories~\cite{Alles:1996nm,Alles:1997qe,DelDebbio:2004ns,DelDebbio:2002xa,DElia:2003zne,Lucini:2004yh,Giusti:2007tu,Vicari:2008jw,Panagopoulos:2011rb,Bonati:2013tt,Ce:2015qha,Ce:2016awn,Berkowitz:2015aua,Borsanyi:2015cka,Bonati:2015sqt,Bonati:2016tvi,
Bonati:2018rfg, Bonati:2019kmf}, with particular focus on the large-$N$ limit
of these models, due to its relation with the Witten--Veneziano
mechanism~\cite{Witten:1979vv,
Veneziano:1979ec,Kawarabayashi:1980dp,Witten:1980sp}.\\
Concerning two-dimensional models, the $\theta$-dependence of $2d$
$\mathrm{CP}^{N-1}$ models is amenable to be exactly computed using analytical
methods in the large-$N$ limit up to the next-to-leading-order in the $1/N$
expansion~\cite{DAdda:1978vbw, Campostrini:1991kv,
DelDebbio:2006yuf,Rossi:2016uce, Bonati:2016tvi}, and such predictions have been
verified to be supported by numerical
evidence~\cite{Vicari:1992jy,Bonanno:2018xtd,Berni:2019bch,Bonati:2019olo,Bonati:2019ylr}.
The $\theta$-dependence of $2d$ $\mathrm{CP}^{N-1}$ models has also been
studied numerically~\cite{Berni:2020ebn,Bonanno:2022dru} in the limit $N\to2$,
as these theories reduce to the $2d$ O(3) non-linear $\sigma$ model in this
limit (see also~\cite{Nogradi:2012dj,Alles:2014tta}), whose $\theta$-dependence
is theoretically interesting due to its connection with the Haldane conjecture.
Finally, also the $\theta$-dependence of $2d$ $\mathrm{U}(N)$ Yang--Mills
theories can be exactly computed analytically~\cite{Kovacs:1995nn, Bonati:2019ylr, Bonati:2019olo}.

The $\theta$-dependence of the spectrum of the theory has received comparatively
less attention, with only one exploratory lattice study~\cite{DelDebbio:2006yuf} present
in the literature. 
Our main goal is to bridge that gap. 
In particular, we report on our study~\cite{Bonanno:2024ggk} of 
the $\theta$-dependence of the spectrum of glueballs and fluxtubes of pure-gauge
$\SU(N)$ models in $4d$.
The results that we have obtained constitute a substantial improvement over
the previously available results. This could be achieved with the combination 
of the imaginary-$\theta$ method and of the Parallel Tempering on Boundary
Conditions (PTBC) algorithm. The former enables us to improve the signal-to-noise
ratio compared to the standard Taylor expansion approach, and the
latter enabled us to avoid the freezing of topology, especially at large-$N$.

This proceeding is organized as follows. In Sec.~\ref{sec:setup} we describe
our numerical setup. In Sec.~\ref{sec:res} we present the main results
of~\cite{Bonanno:2024ggk}. Finally, in Sec.~\ref{sec:conclu} we draw our
conclusions.

\section{Numerical Setup}\label{sec:setup}

Direct lattice simulations of the Yang--Mills theory at non-zero values of
$\theta$ are hindered by the infamous sign problem, as the topological term is
purely imaginary, thus yielding a complex action. A popular technique to bypass
the
sign-problem is to resort to simulations of imaginary-values of $\theta_{_\I}
\equiv \mathrm{i}\theta$, characterized by a purely-real action.
Assuming analyticity around $\theta=0$ it is possible to use analytic
continuation and infer the dependence on the real parameter $\theta$ from the
observed dependence on the imaginary one $\theta_{_\I}$, at least for small
enough values of $\theta$. The imaginary-$\theta$ method has been shown to be extremely effective in
improving the signal-to-noise-ratio of the higher-order coefficients in the
$\theta$-expansion of several quantities, compared to simulations at $\theta=0$
alone~\cite{Bhanot:1984rx, Azcoiti:2002vk, Alles:2007br,
Imachi:2006qq,Aoki:2008gv,Panagopoulos:2011rb, Alles:2014tta, DElia:2012pvq,
DElia:2013uaf,Bonati:2015sqt, Bonati:2016tvi, Bonati:2018rfg,
Bonati:2019kmf,Bonanno:2018xtd, Berni:2019bch, Bonanno:2020hht,
Bonanno:2023hhp}.
Our lattice action at non-zero imaginary-$\theta$ reads:
\beq
\mathcal{S}_\L(\theta_\L) = \mathcal{S}_{_\W} + \theta_\L Q_{\clov},
\eeq
where $S_w$ is the standard Wilson plaquette action,
\beq
\mathcal{S}_{_\W} =
-\frac{\beta}{2 N}\sum_{x}\sum_{\nu>\mu} 
\Tr\left[\mathcal{P}_{\mu\nu}(x)\right]~,
\eeq
with $\beta$ the bare gauge coupling, $\mathcal{P}_{\mu\nu}(x)$ the product of
gauge links around an elementary plaquette,
\beq
\mathcal{P}_{\mu\nu}(x) &=& U_\mu(x) U_\nu(x+a\hat{\mu}) U_\mu^\dag(x+a\hat{\nu}) U_\nu^\dag(x),
\eeq
based in the site $x$ of the lattice and lying in the $(\mu,\nu)$ plane,
and $U_\mu(x) \in \SU(N)$ are gauge link variables. 

For the lattice
topological charge, we employed the standard clover discretization,
\beq
Q_{\clov} = 
\frac{1}{2^9 \pi^2} 
\sum_{\mu\nu\rho\sigma 
= \pm1}^{\pm4} \varepsilon_{\mu\nu\rho\sigma} 
\Tr\left[ \mathcal{P}_{\mu\nu}(x) \mathcal{P}_{\rho\sigma}(x) \right]~.
\eeq
$Q_{\clov}$ is not integer-valued on the lattice and is related to the continuum 
topological charge via a finite renormalization~\cite{Campostrini:1988cy},
$Q_{\clov}=Z_{_Q}Q$, where $Z_{_Q}(\beta)<1$ tends to $1$ only in the continuum
limit ($\beta\to\infty$). As a result, the lattice parameter $\theta_\L$ is thus
related to the physical one by: $\theta = \mathrm{i} Z_{_Q} \theta_\L$.

Obtaining the renormalization constant $Z_{_Q}$ requires the calculation of the lattice topological charge. This can be computed by means of smoothing. To
this end one can adopt any of the various algorithms that have been proposed in
the literature, such as
cooling~\cite{Berg:1981nw,Iwasaki:1983bv,Itoh:1984pr,Teper:1985rb,
Ilgenfritz:1985dz,Campostrini:1989dh,Alles:2000sc}, smearing~\cite{APE:1987ehd,
Morningstar:2003gk}, or gradient flow~\cite{Luscher:2009eq,Luscher:2010iy,
Luscher:2011bx}, as they have been shown to be all numerically equivalent when
matched to one another~\cite{Alles:2000sc, Bonati:2014tqa,Alexandrou:2015yba}.
In this work, we determine the lattice topological charge and the renormalization factor $Z_{_Q}$ via cooling as follows~\cite{Panagopoulos:2011rb}:
\beq
Z_Q = \frac{\braket{Q_{_\I} Q_\clov}}{\braket{Q_{_\I}^2}}, \quad Q_{_\I} = \mathrm{round}\left\{\alpha \, Q_\clov^{(\cool)}\right\},
\eeq
with $Q_{_\I}$ the lattice integer-valued topological
charge~\cite{DelDebbio:2002xa}, $Q_\clov^{(\cool)}$ the clover-discretized charge computed after 20
cooling steps, and $\alpha$ the numerical value minimizing
\beq
\left\langle\left(
\alpha Q_\clov^{(\cool)} - \mathrm{round}\left\{\alpha \,
Q_\clov^{(\cool)}\right\}
\right)^2\right\rangle,
\quad 1 < \alpha < 2.
\eeq

We performed simulations for $N=3$ and $N=6$ on hypercubic lattices $L^4$ and
several values of $\beta$. For the $N=3$ runs, we adopted the standard 4:1
mixture of over-relaxation~\cite{Creutz:1987xi} and
heat-bath~\cite{Creutz:1980zw, Kennedy:1985nu} algorithms. For $N=6$ runs we
instead adopted the Parallel Tempering on Boundary Conditions (PTBC)
algorithm~\cite{Hasenbusch:2017unr} to overcome the severe critical slowing down
experienced at large-$N$ and close to the continuum limit by topological
modes~\cite{Alles:1996vn,DelDebbio:2004xh, Schaefer:2010hu}, known as
\emph{topological freezing}. The PTBC algorithm has been extensively applied
both in $2d$ models~\cite{Hasenbusch:2017unr,Berni:2019bch, Bonanno:2022hmz} and
in $4d$ gauge theories, both with~\cite{Bonanno:2024zyn} and
without~\cite{Bonanno:2020hht,Bonanno:2022yjr,DasilvaGolan:2023cjw,Bonanno:2023hhp,Bonanno:2024nba}
dynamical fermions. This algorithm consists in simulating several
replicas of the lattice, all differing from each other by just for the boundary
conditions imposed
on a small sub-region of the lattice, known as the defect. Boundary conditions
are taken
to be periodic everywhere but on the defect where, on each replica, they are
chosen to interpolate between periodic and open. The state of each 
replica is evolved
with a standard local Monte Carlo updating algorithm except for swaps between
gauge
configurations of the different replicas, that are proposed and stochastically
accepted/rejected via a standard Metropolis step. The calculation of observables
is always performed on the replica with periodic boundary conditions. This
algorithm enables us to enjoy the fast decorrelation of topological modes
achieved with open boundaries (which is transferred towards the periodic replica
by the swaps) and, at the same time, to avoid the systematic effects introduced
with open boundary conditions.

We determine the $\SU(N)$ spectrum with the variational method. An appropriate
variational basis of zero-momentum projected operators $\left\{O_i(t)\right\}$
was defined for each symmetry channel of interest.
In particular, the operators were path ordered products of link 
variables along space-like paths of various shapes and sizes, and at
different levels of blocking and
smearing~\cite{BERG1983109,APE:1987ehd,TEPER1987345,
Teper:1998te,Lucini:2001ej,Lucini:2004my,Blossier:2009kd, Lucini:2010nv,
Bennett:2020qtj,Athenodorou:2020ani, Athenodorou:2021qvs}. Their correlator
was computed on the generated ensembles, 
\beq
C_{ij}(t) = \braket{O_i(t)O_j(0)}
\eeq
and then a Generalized EigenValue Problem (GEVP), $C_{ij}(t) v_j = \lambda(t,t_0)C_{ij}(t_0)v_j$, was solved. This enabled us to find the
optimal combination of operators for each symmetry channel, and to define
the corresponding correlation function,
\beq
\overline{C}(t) = C_{ij}(t)\overline{v}_j,
\eeq
with $\overline{v}$ the eigenvector related to the largest eigenvalue
$\lambda(t,t_0)$. The GEVP was always solved for $t_0=a$, and we checked in a
few cases that $t_0=2a$ gave compatible results. 
The ground state mass was then obtained via a best fit of 
$A \left[ \exp(-mt)+\exp(-m(La-t)) \right]$
to the correlator, using $m$ and $A$ as fitting parameters,
over a range of $t$ where the effective mass, defined as follows,
\beq
a m_{_{\mathrm{eff}}}(t) = 
- \log\left[\frac{\overline{C}(t+a)}{\overline{C}(t)}\right]~,
\eeq
exhibited a plateaux.

For the mass gap of the theory, we considered a variational basis made of 4-, 6-
and 8-link operators in the $\mathrm{A}_1$
representation of the octahedral group. For the torelon mass we used a
variational basis made of products of fat-links winding around the time
direction once. We then extracted the string tension
using~\cite{deForcrand:1984wzs} (with $L$ the lattice size in lattice units):
\beq
a^2\sigma = \frac{a m_{\tor}}{L} + \frac{\pi}{3 L^2}.
\eeq

\section{Results}\label{sec:res}

\begin{figure}[!t]
\centering
\includegraphics[scale=0.23]{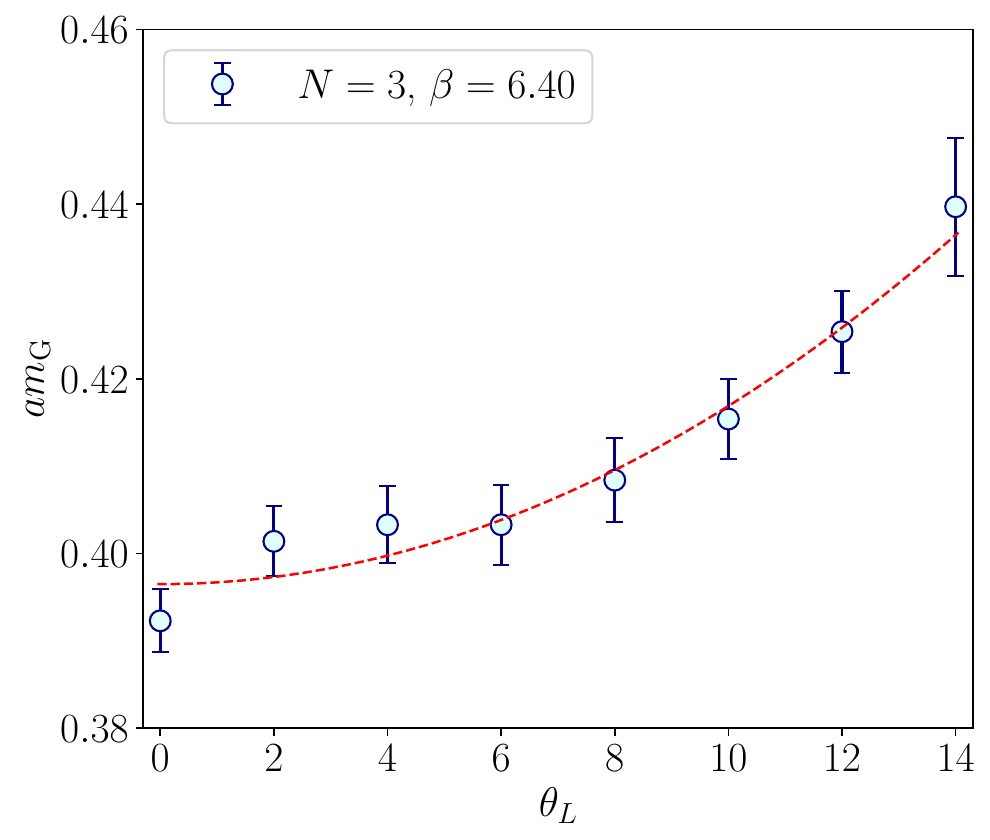}
\includegraphics[scale=0.23]{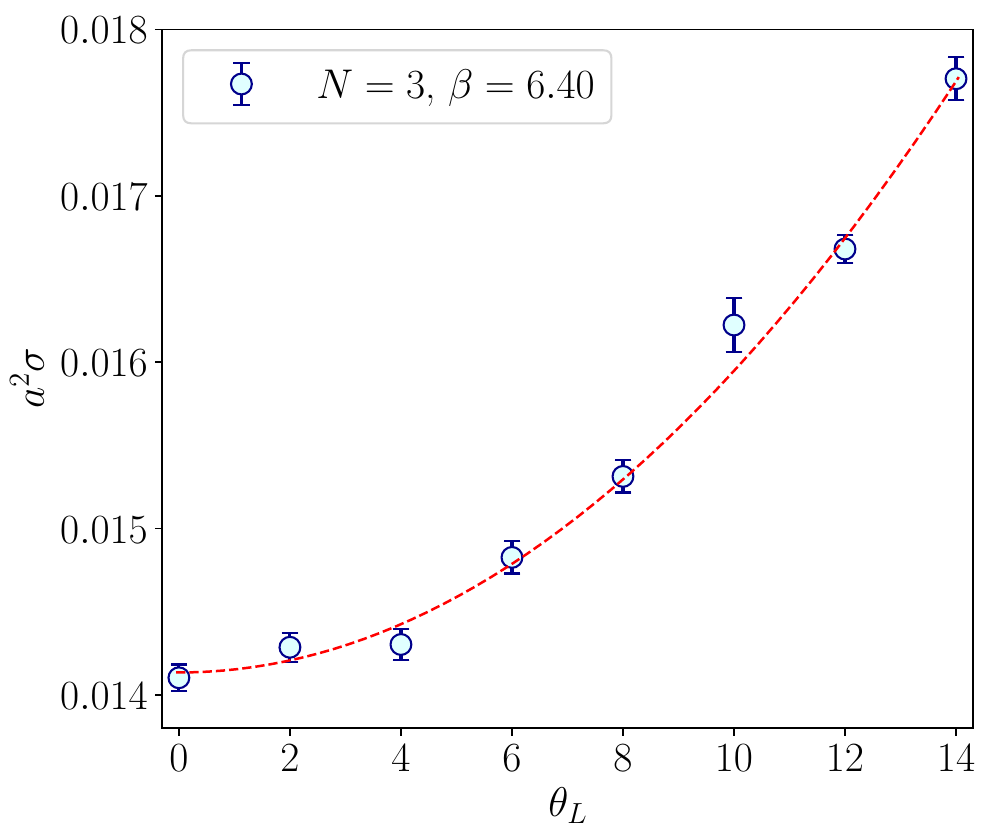}
\includegraphics[scale=0.23]{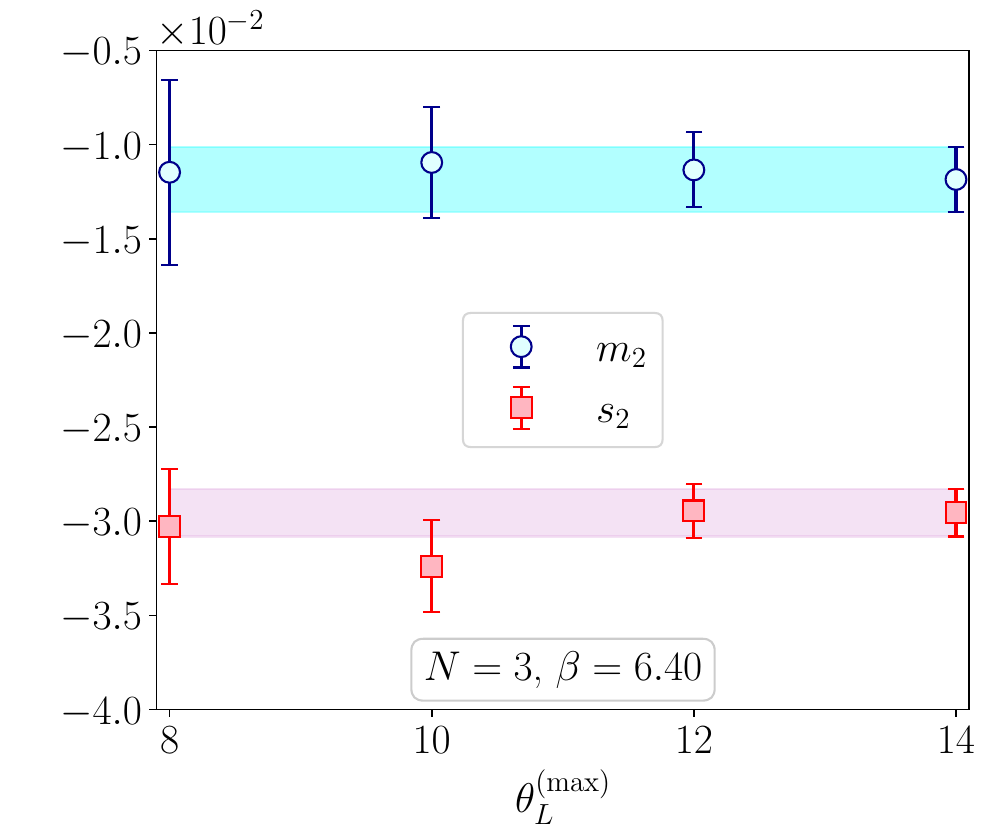}
	
\includegraphics[scale=0.23]{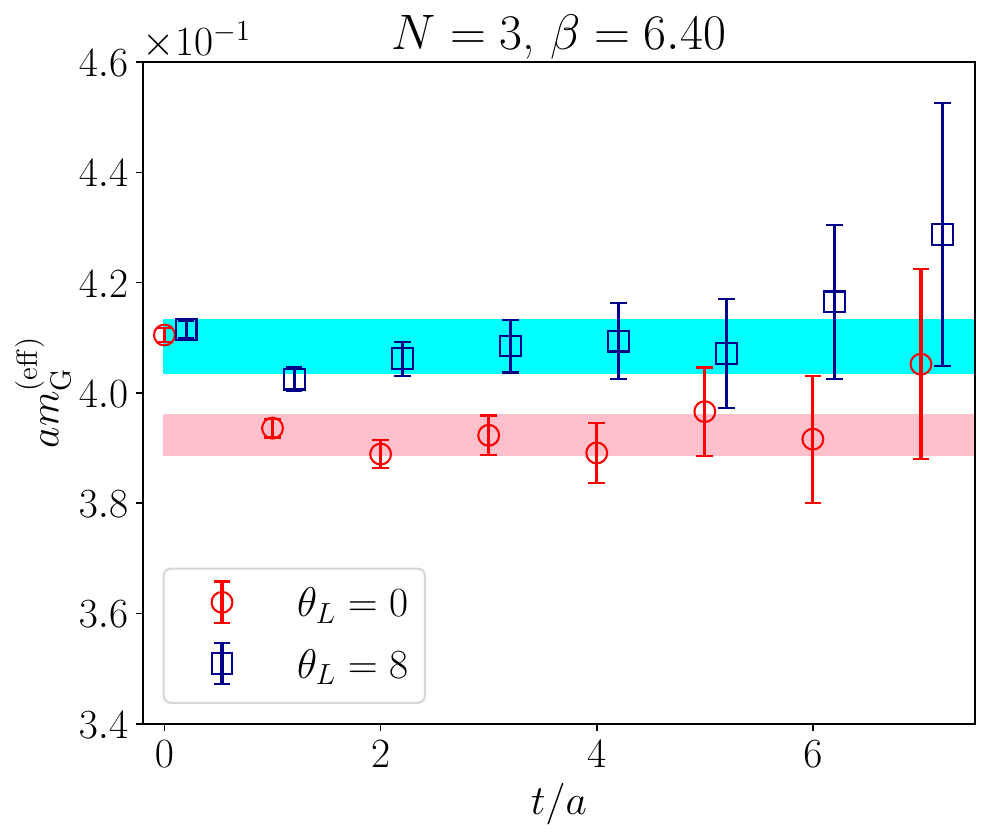}
\includegraphics[scale=0.23]{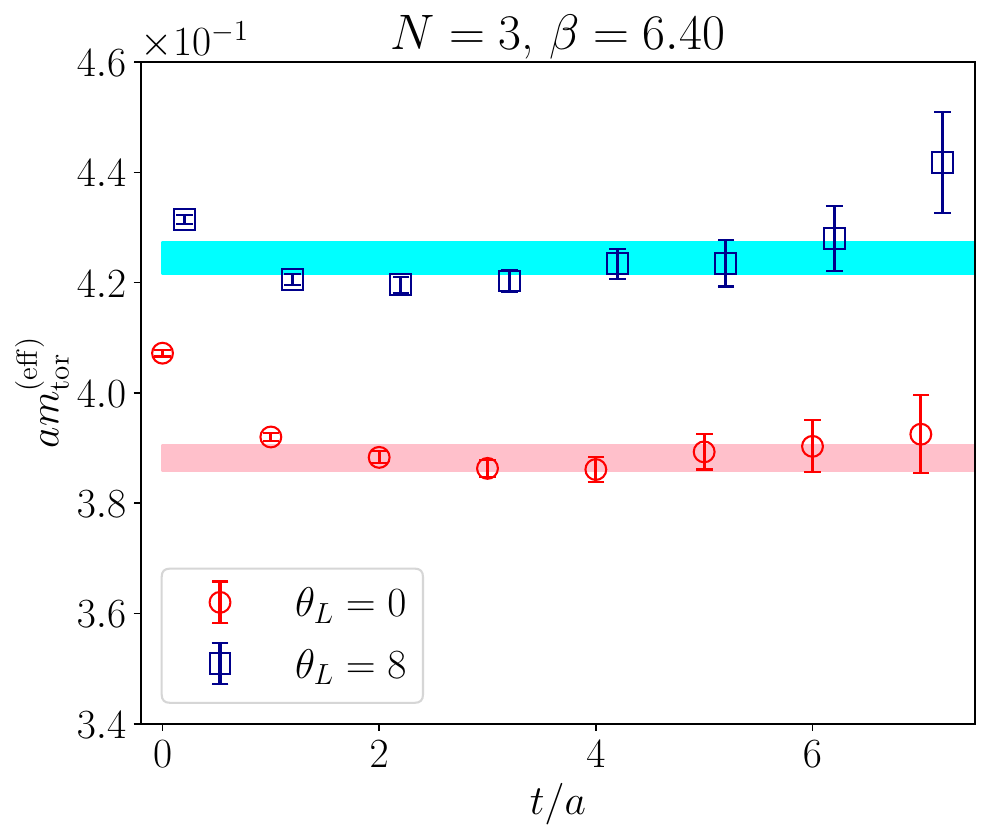}
\caption{Results for $N=3$ with $\beta=6.40$ (finest lattice spacing explored).}
\label{fig:ex_fit_theta_N3}
\end{figure}

We determined the mass gap of the theory (i.e., the mass of the lightest
glueball state) and the string tension in lattice units for several values of
$\beta$ and $\theta_{\L}$. 
The $\theta$ dependence was parameterized by Taylor expanding in
up to the next-to-leading order, around $\theta=0$,
\beq
am_{\G}(\theta) &=& am_{\zpp} \left[1 + m_2 \theta^2 + \mathcal{O}(\theta^2)\right],\\
a^2\sigma(\theta) &=& a^2\sigma \left[1 + s_2 \theta^2 + \mathcal{O}(\theta^2)\right]~,
\eeq
where the well-established fact that at $\theta=0$ the mass of the $0^{++}$
glueball is the lightest one in pure-Yang--Mills theories~\cite{Lucini:2004my,
Athenodorou:2020ani, Athenodorou:2021qvs,Vadacchino:2023vnc}) was used. 
Using analyticity, and the renormalization property of the lattice
imaginary-$\theta$ parameter, one simply has:
\beq
am_{\G}(\theta_{\L}) &=& am_{\zpp}\left[ 1 - m_2 Z_{_Q}^2\theta_\L^2 + \mathcal{O}(\theta_\L^2)\right],\\
a^2\sigma(\theta_{\L}) &=& a^2\sigma\left[ 1 - s_2 Z_{_Q}^2\theta_\L^2 + \mathcal{O}(\theta_\L^2) \right].
\eeq
We thus performed a best fit of our determinations of $m_{\G}(\theta_{\L})$ and
$\sigma(\theta_{\L})$ as a function of $\theta_\L$ at fixed $\beta$ and
determined the quantities $m_2 Z_{_Q}^2$ and $s_2 Z_{_Q}^2$. 
Since the value of $Z_{_Q}(\beta)$ was known 
for each $\beta$, we could obtain $m_2$ and $s_2$. Our
results for  $m_2$ and $s_2$ at $N=3$ are displayed in
Fig.~\ref{fig:ex_fit_theta_N3}. They are very stable as a function of the 
fitting range, signaling that we are insensitive to the effects of 
higher-order corrections. We also display a few examples of plateaux of
effective masses.

\begin{figure}[!t]
\centering
\includegraphics[scale=0.35]{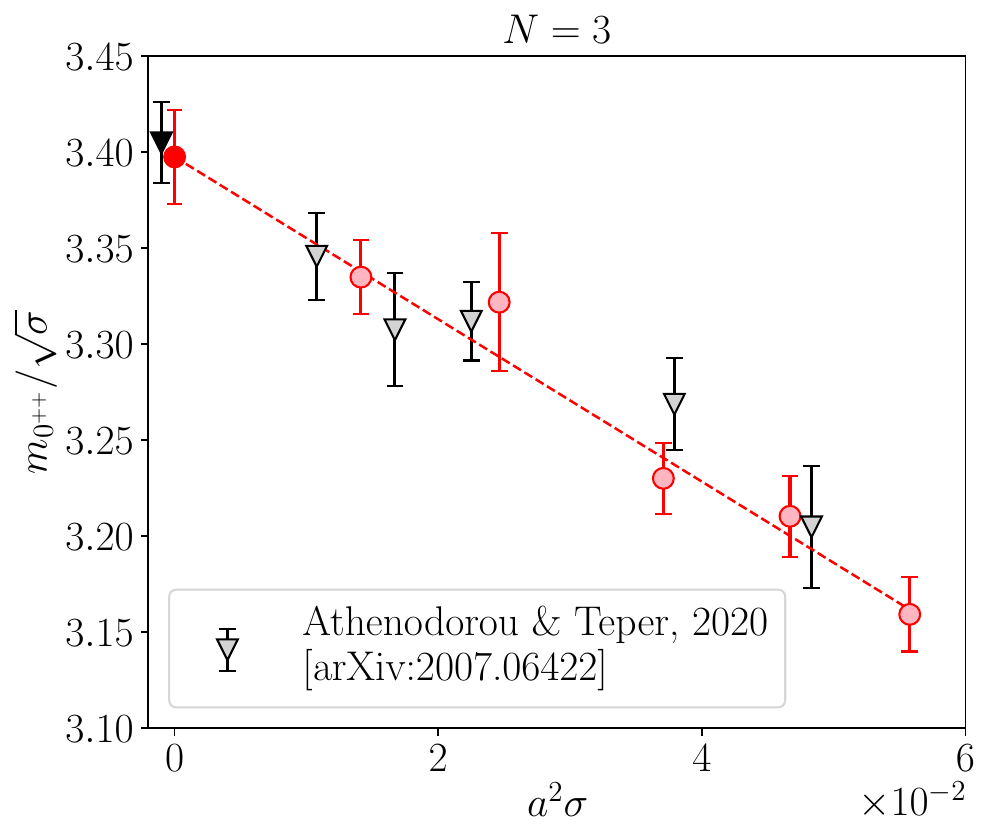}
\includegraphics[scale=0.35]{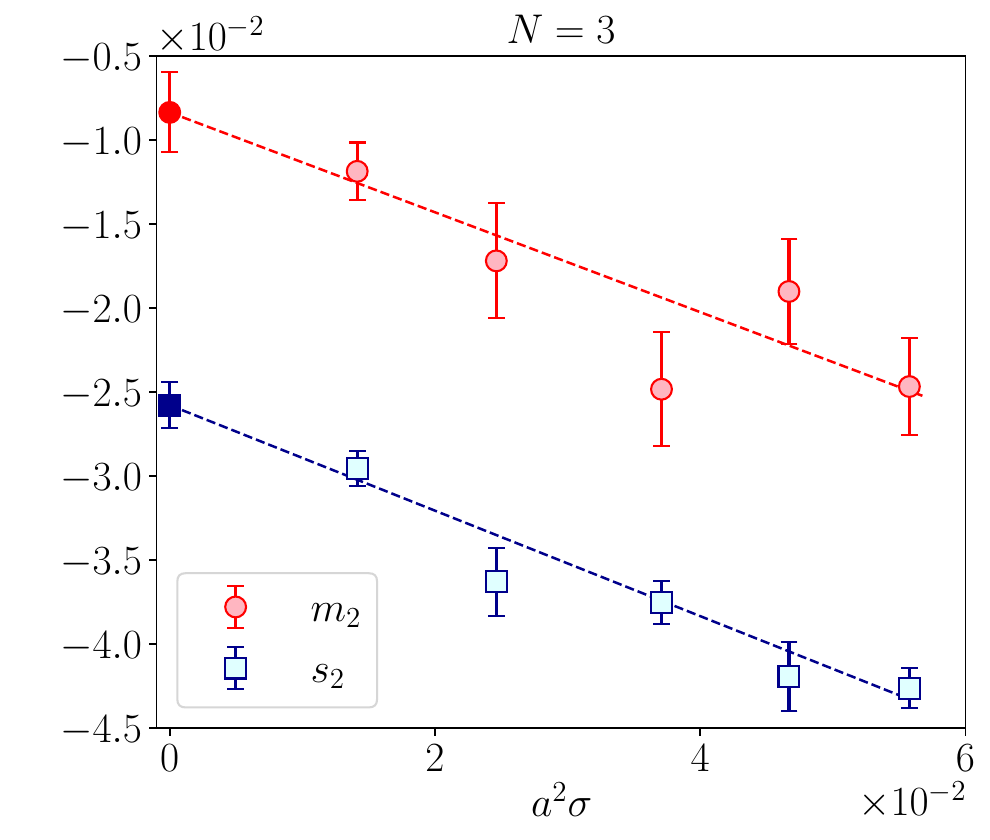}
\caption{Continuum limit of our $N=3$ results for $m_2$ and $s_2$. We also computed the continuum limit of
$m_{\zpp}/\sqrt{\sigma}$, which is in perfect agreement with the previous result of~\cite{Athenodorou:2020ani}.}
\label{fig:ex_cont_limit_N3}
\end{figure}

In Fig.~\ref{fig:ex_cont_limit_N3} we display the continuum extrapolation of our
results for $m_2$ and $s_2$. We also extrapolated towards the continuum limit
the dimensionless ratio $m_{\zpp}/\sqrt{\sigma}$ in order to verify that our
result is in agreement with the previous determination provided in
Ref.~\cite{Athenodorou:2020ani}.

\begin{figure}[!t]
\centering
\includegraphics[scale=0.35]{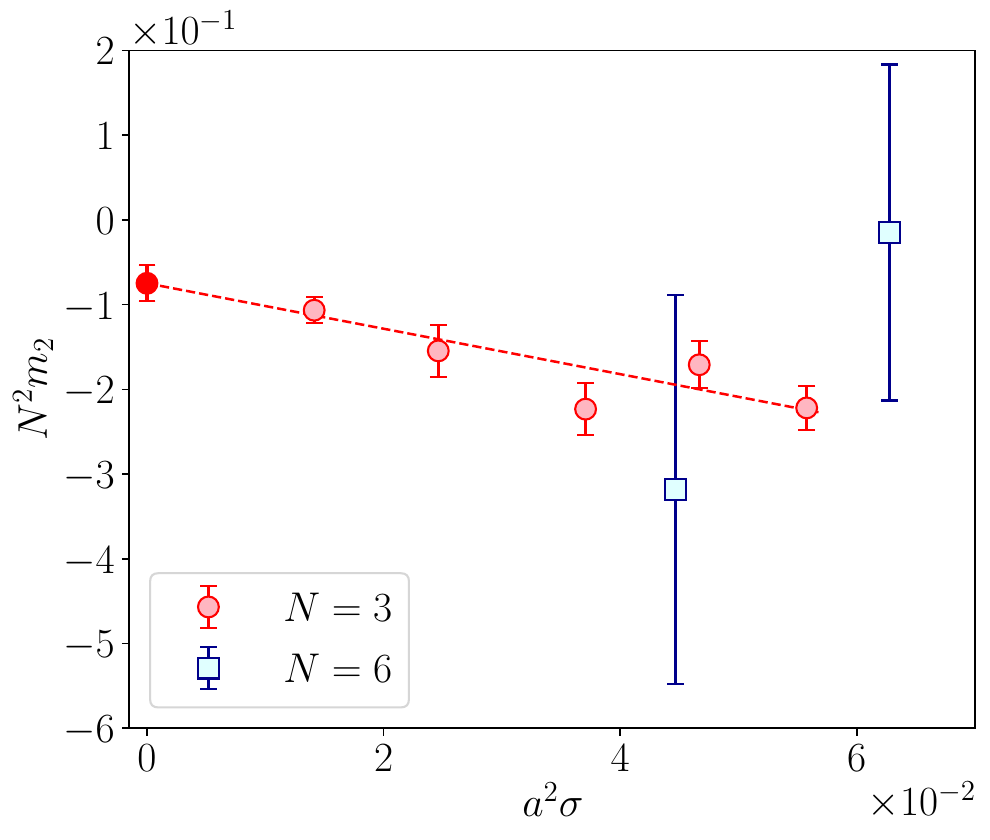}
\includegraphics[scale=0.35]{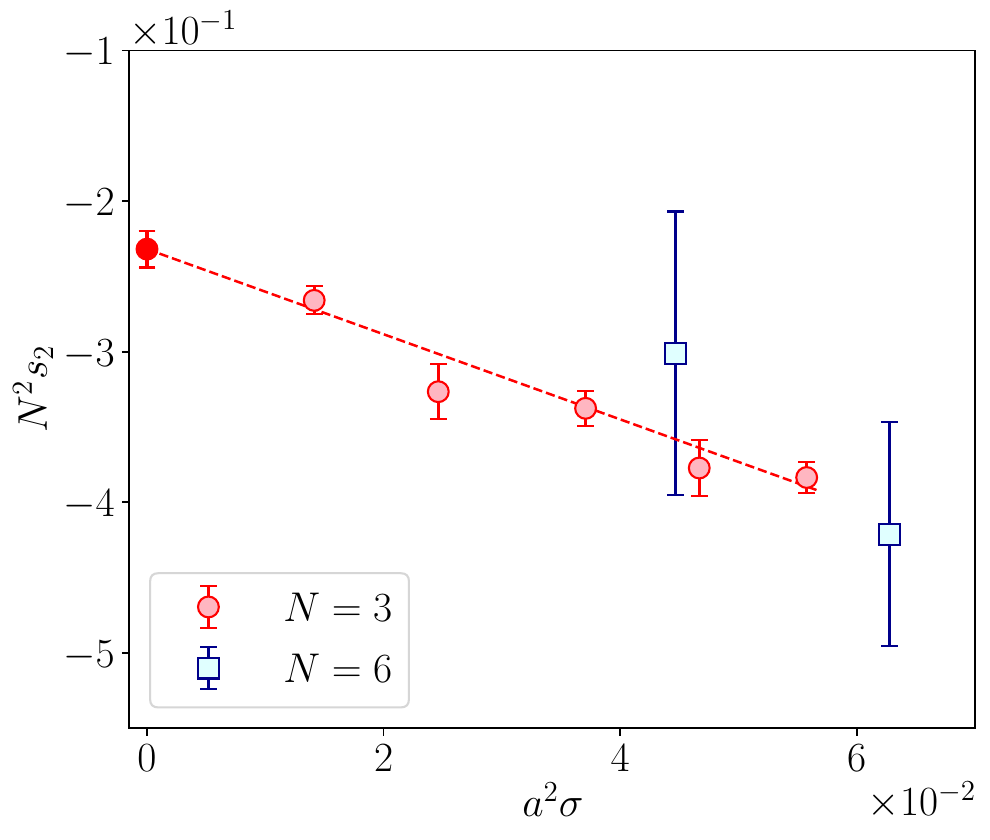}
\caption{Continuum scaling of $N^2 m_2$ and $N^2 s_2$ for $N=3$ and $N=6$.} 
\label{fig:ex_cont_limit_N6}
\end{figure}

These are our final determinations for $N=3$, in the continuum limit,
\beq
\label{eq:msN3}
\frac{m_{\zpp}}{\sqrt{\sigma}} = 3.398(25), \qquad \text{(continuum limit, $N=3$)},\\
\label{eq:m2N3}
m_2= -0.0083(23), \qquad \text{(continuum limit, $N=3$)},\\
\label{eq:s2N3}
s_2= -0.0258(14), \qquad \text{(continuum limit, $N=3$)}.
\eeq

Concerning $N=6$, we obtained results only for two fairly fine lattice spacings, thus we cannot perform a continuum limit of these data alone. However, our data allowed for a first quantitative check of the following expected large-$N$ scaling:
\beq
m_2 = \frac{\overline{m}_2}{N^2} + \mathcal{O}\left(\frac{1}{N^4}\right),\quad
s_2 = \frac{\overline{s}_2}{N^2} + \mathcal{O}\left(\frac{1}{N^4}\right).
\eeq
Our $N=3$ and 6 results are perfectly compatible with this expected scaling, see Fig.~\ref{fig:ex_cont_limit_N6}. Thus, in the end we can quote the following estimates:
\begin{equation}\label{eq:bar_res}
\overline{s}_2 \simeq -0.23(1)\ ,\quad
\overline{m}_2 \simeq -0.075(20)\ .
\end{equation}

\section{Conclusions}\label{sec:conclu}

The present manuscript reports on the main finding of~\cite{Bonanno:2024ggk}. We
have studied the $\theta$-dependence of the $\SU(N)$ spectrum, focusing on the
mass gap of the theory and on the string tension for $N=3$ and $N=6$.

For $\SU(3)$, in the continuum limit we found:
\beq
\frac{m_{\G}}{\sqrt{\sigma}}\Bigg\vert_{\theta=0} 
= \frac{m_{\zpp}}{\sqrt{\sigma}} 
= 3.398(25)~,\quad
m_2 = -0.0083(23),\quad
s_2 = -0.0258(14), 
\eeq
where 
\beq
m_2\equiv \frac{1}{2m_{\zpp}}\times
\frac{\mathrm{d}m_{\G}(\theta)}{\mathrm{d}\theta^2}\Bigg\vert_{\theta\,=\,0},\quad
s_2\equiv \frac{1}{2\sigma}\times
\frac{\mathrm{d}\sigma(\theta)}{\mathrm{d}\theta^2}\Bigg\vert_{\theta\,=\,0}.
\eeq
The results obtained at $N=6$ also enabled us to check that 
the expected large-$N$ scaling was realized and to estimate:
\beq
m_2(N) \simeq 
\frac{-0.075(20)}{N^2}+\mathcal{O}\left(\frac{1}{N^4}\right),\quad
s_2(N) \simeq 
\frac{-0.23(1)}{N^2}+\mathcal{O}\left(\frac{1}{N^4}\right).
\eeq

Note that for $N=3$, the value of $m_2$ is very close to $s_2/2$.
As a result, the coefficient $g_2$ of the order $\mathcal{O}(\theta^2)$ 
correction to the dimensionless ratio $m_\G/\sqrt{\sigma}$,
\beq
\left(\frac{m_\G}{\sqrt{\sigma}}\right)(\theta) =
\frac{m_\zpp}{\sqrt{\sigma}}\left(1+g_2\theta^2+\mathcal{O}(\theta^4)\right),
\eeq
is compatible with zero within two standard deviations,
\beq
g_2 = m_2 - \frac{s_2}{2} = 0.0046(24)~.
\eeq
We are not aware of any general theoretical argument that would dictate the
$\theta$-independence of this dimensionless ratio. Actually, we can 
provide a counter-example to this. In Refs.~\cite{DElia:2012pvq,DElia:2013uaf,
Bonanno:2023hhp}, the $\theta$-dependence of the $\SU(N)$ deconfinement critical
temperature $T_c$ was investigated for $N\ge 3$ (for a first SU(2) study
see~\cite{Yamada:2024pjy}). It was concluded that if
\beq
T_c(\theta) = T_c[1 - R \theta^2 + \mathcal{O}(\theta^4)]~,
\eeq
then 
\beq
R(N=3) = 0.0178(5)~,\quad 
R(N)= \frac{0.159(4)}{N^2}+\mathcal{O}\left(\frac{1}{N^4}\right) \quad (N>3)~.
\eeq
Using our result for $s_2$, we would then find,
\beq
\frac{T_c(\theta)}{\sqrt{\sigma(\theta)}} = 
\frac{T_c}{\sqrt{\sigma}}[1 -t_2\theta^2+ \mathcal{O}(\theta^4)]~,
\eeq
and the coefficient $t_2$ would then be non-vanishing, 
\beq
t_2 = R+\frac{s_2}{2} = 0.0049(9)~.
\eeq

It is interesting to observe that the $\theta$-dependence of $m_{\G}$ and $\sigma$ in large-$N$ Yang--Mills theories has been also addressed within holographic models in~\cite{Bigazzi:2015bna}, providing the following prediction:
\beq
\frac{s_2}{m_2} = \frac{\overline{s}_2}{\overline{m}_2} = 4, \qquad \qquad
\text{(holography),}
\eeq
which nicely agrees with our lattice result
\beq
\frac{s_2}{m_2} = \frac{\overline{s}_2}{\overline{m}_2} = 3.07(82), \qquad \qquad \text{(lattice).}
\eeq
On the other hand, Ref.~\cite{Bigazzi:2015bna} predicts that the ratio $T_c(\theta)/m_{\G}(\theta)$ is $\theta$-independent at $\mathcal{O}(\theta^2)$, i.e., $R/m_2 = -1$, in contrast with our lattice result $R/m_2 = -2.14(57)$ indicating that $T_c(\theta)/m_{\G}(\theta)$ has a non-trivial $\theta$-dependence already at leading order in $\theta$.

Finally, we recall the interesting role played by $m_2$, identified
in~\cite{Brower:2003yx, Aoki:2007ka} in the estimation of the systematic
error introduced in lattice spectra calculations performed at fixed
topological sector. At a fixed value $Q$ of the topological charge, 
the following approximate relation holds for the mass $M$ of 
any bound state,
\beq\label{eq:Qbias}
\frac{M^{(Q)} - M}{M} \approx \frac{M_2}{2 \chi V}~,
\eeq
where $\chi=\braket{Q^2}/V$ is the topological susceptibility and $M_2$ 
is the order $\mathcal{O}(\theta)^2$ coefficient of
$M(\theta)=M[1+M_2\theta^2+\mathcal{O}(\theta^4)]$. For $N=3$ the topological
susceptibility is roughly $\chi\simeq (1\text{ fm})^{-4}$. For a standard
lattice volume of $V=(1.5\text{ fm})^4$, one has the following bound on the systematic error on the numerical estimation of the lightest $0^{++}$ glueball:
\beq
\frac{\Delta m_{\zpp}}{m_{\zpp}}\Bigg\vert_{N=3} \approx \frac{m_2}{2\chi V} \approx -0.08\%.
\eeq
This bound will become even more favorable at larger value of $N$, since $\chi$ does not change appreciably with $N$, while $m_2 \sim \mathcal{O}(1/N^2)$.

\acknowledgments
The work of C.~Bonanno is supported by the Spanish Research Agency (Agencia Estatal de Investigación) through the grant IFT Centro de Excelencia Severo Ochoa CEX2020- 001007-S and, partially, by grant PID2021-127526NB-I00, both funded by MCIN/AEI/10.13039/ 501100011033. C.~Bonanno also acknowledges support from the project H2020-MSCAITN-2018-813942 (EuroPLEx) and the EU Horizon 2020 research and innovation programme, STRONG-2020 project, under
grant agreement No 824093. The work of D.~Vadacchino is supported by STFC under Consolidated Grant No.~ST/X000680/1. Numerical calculations have been performed on the \texttt{Galileo100} machine at Cineca, based on the project IscrB\_ITDGBM, on the \texttt{Marconi} machine at Cineca based on the agreement between INFN and Cineca (under project INF22\_npqcd), and on the Plymouth University cluster.

\bibliographystyle{JHEP}
\bibliography{biblio}

\end{document}